\documentclass[aps,prr,twocolumn,showpacs,superscriptaddress,10pt]{revtex4-1}

\usepackage[colorlinks ,linkcolor=magenta,anchorcolor=green,citecolor=blue,urlcolor=blue]{hyperref}
\usepackage{amsmath,amsfonts,amssymb,mathrsfs}
\usepackage{bm,bbm}
\usepackage{graphicx}
\usepackage{epstopdf}
\usepackage[dvipsnames]{xcolor}
\usepackage{verbatim}
\usepackage[normalem]{ulem}

\usepackage[T1]{fontenc}
\usepackage{tgtermes}

\newcommand{\appsection}[1]{\section{\uppercase{#1}}}

\begin{document}

\title{Trion states and quantum criticality of attractive SU(3) Dirac fermions}
\author{Han Xu}
\affiliation{School of Physics and Technology, Wuhan University, Wuhan
430072, China}
\affiliation{Department of Physics, City University of Hong Kong, Tat Chee Avenue, Kowloon, Hong Kong SAR, China, and City University of Hong Kong Shenzhen Research Institute, Shenzhen, Guangdong 518057, China}

\author{Xiang Li}
\affiliation{School of Physics and Technology, Wuhan University, Wuhan
430072, China}

\author{Zhichao Zhou}
\affiliation{School of Physics and Technology, Wuhan University, Wuhan
430072, China}

\author{Xin Wang}
\affiliation{Department of Physics, City University of Hong Kong, Tat Chee Avenue, Kowloon, Hong Kong SAR, China, and City University of Hong Kong Shenzhen Research Institute, Shenzhen, Guangdong 518057, China}

\author{Lei Wang}
\affiliation{Institute of Physics, Chinese Academy of Sciences, Beijing 100190, China}

\author{Congjun Wu}
\affiliation{Department of Physics, School of Science, Westlake University, Hangzhou 310024, Zhejiang, China}
\affiliation{Institute for Theoretical Sciences, Westlake University, Hangzhou 310024, Zhejiang, China}
\affiliation{Key Laboratory for Quantum Materials of Zhejiang Province, School of Science, Westlake University, Hangzhou 310024, China}
\affiliation{Institute of Natural Sciences, Westlake Institute for Advanced Study, Hangzhou 310024, Zhejiang, China}

\author{Yu Wang}
\email{yu.wang@whu.edu.cn}
\affiliation{School of Physics and Technology, Wuhan University, Wuhan
430072, China}

\begin{abstract}
We perform the projector quantum Monte Carlo (QMC) simulation to study the trion formation and quantum phase transition in the half-filled attractive SU(3) Hubbard model on a honeycomb lattice. With increasing attractive Hubbard interaction, our simulations demonstrate a continuous quantum phase transition from the semimetal to charge density wave (CDW) at the critical coupling $U_c/t=-1.52(2)$. The critical exponents $\nu=0.82(3)$ and $\eta=0.58(4)$ determined by the QMC simulation remarkably disagree with those of the $N=3$ chiral Ising universality class suggested by the effective Gross-Neveu-Yukawa (GNY) theory, but coincide with the $N=1$ chiral Ising universality class. In the CDW phase, we show that on-site and off-site trions coexist and the off-site trion forms a local bond state. Our work not only illustrates the formation of off-site trions in two-dimensional Hubbard model, but also raises doubts about the extent of applicability of GNY model on the attractive SU(3) Dirac fermions.
\end{abstract}

\maketitle

\section{Introduction}
Optical traps and lattices loaded with ultracold atoms have become excellent platforms for studying strong correlation physics. %Physical parameters of optical lattices, e.g. lattice structures and interactions, are highly controllable. , which opens an avenue for simulating various Hamiltonians.
Interestingly, since the ultracold alkali and alkaline-earth fermions can carry large hyperfine spins, they provide an opportunity to study SU($N$) ($N > 2$) symmetries that are typical in high energy physics but rare in solids. In recent decades, fermionic models with SU($N$) symmetry have been of great interest to both experimentalists  \cite{Taie2010,DeSalvo2010,Taie2012,Scazza2014,Zhang2014,Pagano2014,Cazalilla2014,Hofrichter2016,  Riegger2018Localized,he2020collective,Song2020Evidence,ozawa2018,Taie2022}
and theorists \cite{Wu2003,Honerkamp2004Ultra,Honerkamp2004BCS,Wu2005a,Gorshkov2010,Yoshida2021Rigorous} in the interdisciplinary context of ultracold atom physics and condensed matter physics.
In particular, the SU($3$) model, as a minimal SU($N$) model beyond SU(2), increasingly arouses interest of researchers because of its striking resemblance to the quark matter \cite{Fodor2002,Aoki2006,Wilczek2007}.
The SU(3) symmetry can be experimentally realized with ultracold fermionic $^{6}\mathrm{Li}$ atoms \cite{Abraham1997,Bartenstein2005,Ottenstein2008,Huckans2009}. Each $^{6}\mathrm{Li}$ atom is in its three lowest hyperfine states, hereinafter referred to as ``colors''. When the three pairwise $s$-wave scattering lengths approach a common negative value, the attractive interactions become SU(3) symmetric \cite{Huckans2009}.

The SU(3) $^{6}\mathrm{Li}$ Fermi gas with tunable interactions is exceptionally appropriate for studying the three-body bound states \cite{Wenz2009Universal,Williams2009Evidence,nakajima2010nonuniversal}. In few-body spin-$1/2$ fermion systems, forming three-body bound states by on-site attractions is rather challenging due to the Pauli exclusion \cite{mattis1984three,rudin1985absence,mattis1986few}, although three-body bound states can be realized with finite-radius interactions \cite{kornilovitch2013stability}.
However, in attractive few-body three-color fermions, due to an additional internal degree of freedom, three-body bound states can be formed by on-site attractions. Specifically, in the attractive SU($3$) Hubbard model with only three fermions on the honeycomb lattice, the three-body bound states have two configurations under different conditions \cite{Pohlmann2013}: when on-site triple occupancy is energetically favorable, the three-body bound state is an on-site trion, which consists of three fermions at one site;
when on-site triple occupancy takes great energy penalty, the three-body bound state is an off-site trion, which consists of two fermions at one site and one fermion at the nearest-neighbor site. In many-body systems the three-body bound states can develop long-range order. The variational \cite{Rapp2007,Rapp2008}, self-energy functional \cite{Inaba2009,inaba2011color} and dynamical mean-field theory (DMFT) \cite{Titvinidze2011,Koga2017} studies of the SU(3) Hubbard models on two-dimensional lattices have predicted a phase transition between the color superfluid and the on-site trion phase, which is reminiscent of the transition between quark superfluid and baryonic phase \cite{Fodor2002,Aoki2006,Wilczek2007}.  In one-dimensional lattices, the density matrix renormalization group (DMRG) studies have observed off-site trion phase either when on-site triple occupancy is prohibited in an attractive SU(3) Hubbard model \cite{Kantian2009} or when the attractions are color-dependent and thus anisotropic in a three-color Hubbard model with SU(3) symmetry breaking \cite{azaria2009three}.

Quantum Monte Carlo (QMC) simulations of the SU(3) Hubbard model have long been absent due to the notorious sign problem. Owning to recent advances in the QMC algorithm \cite{WangL2015,Li2015b,Wei2016,Li2016}, the QMC simulation of the attractive SU(3) Hubbard model can be proved to be sign-problem-free in bipartite lattices at half filling.
In this work, we propose, for the first time, to conduct a projector determinant QMC simulations of the half-filled attractive SU(3) Hubbard model on the honeycomb lattice.
The purpose of this work is twofold: firstly, QMC simulations of the attractive SU(3) Hubbard model can demonstrate the formation of trionic states, and in particular we shall show that the off-site and on-site trions coexist in this two-dimensional Hubbard model while previously off-site trions have only been found in one-dimensional Hubbard model; and secondly, the QMC simulations of the quantum phase transition of SU(3) Dirac fermions can be compared with the Gross-Neveu-Yukawa (GNY) models which is thought to provide a general description for the criticality of Dirac fermions in two spatial dimensions \cite{zerf2017four}.
Given the symmetry breaking patterns, the GNY descriptions are irrelevant to the details of the microscopic Hamiltonian, but solely dependent on the number of fermion colors, $N$.
So far various QMC calculations are consistent with the GNY models. In the spinless Dirac fermions, the transition between semimetal and charge-density-wave (CDW) phases belongs to the $N=1$ chiral Ising universality class \cite{WangL2014,Li2015c}.
In SU(2) Dirac fermions, the semimetal-CDW, semimetal-antiferromagnet and semimetal-superconductor transitions fall into the $N=2$ chiral Ising \cite{Chen2019}, chiral Heisenberg \cite{Parisen2015,Otsuka2016} and chiral XY \cite{Otsuka2018} universality classes, respectively.
%In SU(2) fermion systems, in the Holstein model the semimetal-CDW transition is consistent with the $N=2$ chiral Ising universality class \cite{Chen2019}. In the Hubbard model, the semimetal-AFM transition belongs to the $N=2$ chiral Heisenberg universality class \cite{Parisen2015,Otsuka2016}. In the attractive Hubbard model, the semimetal-superconductor transition, which breaks the U(1) symmetry, belongs to the $N=2$ chiral XY universality class \cite{Otsuka2018}.
In SU(4) Dirac fermions, the transition between semimetal and valence bond solid (VBS) belongs to the $N=4$ chiral XY universality class, due to an emergent U(1) symmetry at the critical point \cite{Zhou2018Mott,Da2022DiracII,Da2022DiracIII}.
Moreover, in SU($N$) Dirac fermions with singlet-bond interactions, the semimetal-VBS transitions fall into the chiral XY universality classes of $N$ fermion colors \cite{li2017fermion}.
However, we shall show that our QMC simulation of the quantum phase transition in the SU(3) Dirac fermions with attractive Hubbard interactions surprisingly conflicts with the $N=3$ GNY model.

%Recently, rigorous results on the ground state properties prove the formation of CDW order on the bipartite lattice, under the condition that the numbers of sites in the two sublattices differ macroscopically \cite{Yoshida2021Rigorous}.
%Nevertheless, the competition between color superfluid and trion states is inconclusive in general and even the configuration of the trion state . Therefore, using unbiased numerical approaches to obtain controlled and systematic predictions is in high demand.

%%%%%%%%%%%%%%%%%%%%%%%%%%%%%%%%%%%%%%%%%%
\section{Model}
%\label{sect:model}
At half filling, the attractive SU(3) Hubbard model is defined by the lattice Hamiltonian:
\begin{equation}
    H = -t\sum_{\langle{ij}\rangle,\alpha} (c^{\dagger}_{i\alpha}c_{j\alpha}+\mathrm{H.c.}) + U\sum_{i,\alpha<\beta}{(n_{i\alpha}-\frac{1}{2})(n_{i\beta}-\frac{1}{2})},
    \label{eq:hubbard}
\end{equation}
where $\langle{ij}\rangle$ denotes the nearest-neighbor sites on a honeycomb lattice and $\alpha,\beta$ are the color indices running from $1$ to $3$.
$U<0$ describes the on-site attractive Hubbard interaction, and $n_{i\alpha}=c^{\dagger}_{i\alpha}c_{i\alpha}$ is the particle number operator for color $\alpha$ at site $i$.
The nearest-neighbor hopping amplitude $t=1$ is set as the energy unit in our simulations.

\begin{figure}[tb]
    \includegraphics[width=0.5\linewidth]{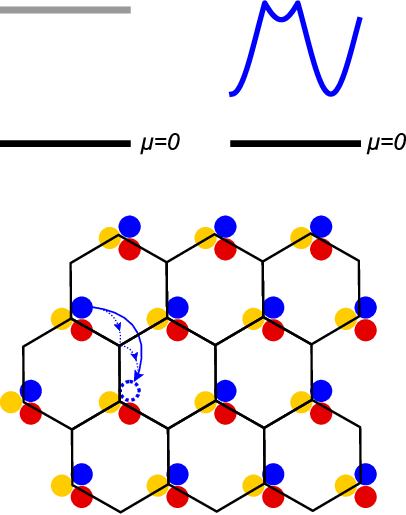}
    \caption{
        Upper panel: The energy penalty of adding a hole to the half-filled system in the atomic limit (upper left) and at strong coupling (upper right). Lower panel: The sketch of a hole hopping process in the background of on-site trions.}
    \label{fig:mottgap}
\end{figure}

In the atomic limit $t/U=0$, on-site trions are randomly formed by three fermions on a lattice point in the half-filled model.
When the fermion hopping is turned on, the charge fluctuations induce the CDW order \cite{Honerkamp2004Ultra}.
As shown in Fig.~\ref{fig:mottgap}, the energy penalty for adding an extra fermion (hole) to the half-filled system is $-U$ in the atomic limit. At strong coupling the extra fermion (hole) can hop on a triangular lattice via second-order perturbation, which expands the energy level into an energy band. Then the energy gap is $\Delta_\mathrm{at}\approx-U-W$ where $W=-\frac{3t^2}{U}$.
At the Mott transition point, the energy gap vanishes, and then the critical coupling strength is estimated as $U^\mathrm{at}_c/t \approx -\sqrt{3}$ (see Appendix~\ref{app:largeUgap}).
% \bibnote[MySuppLink]{See Supplemental Material at URL, which includes Refs.~[], for details}

%-----------------------------------------------------------
%-----------------------------------------------------------

The determinant formalism of the projector QMC method \cite{Assaad2008} will be employed to simulate the attractive SU(3) Hubbard model, which is sign-problem-free at half filling when decomposing the Hubbard interaction into the on-site color-flip channel. Implementation details of the algorithm are elaborated in Appendix~\ref{app:HS}.

%We choose the lattice sizes $L=3,6,9,12$ to touch the Dirac points in the reciprocal lattice and impose the periodic boundary condition along the $x$ and $y$ directions reflecting the translational symmetry. The wave function of the noninteracting SU(3) Dirac fermions with the anti-periodic boundary condition is chosen as the trial wave function. The projection time $\beta=\frac{8}{3}L$ and Suzuki-Trotter discretization $\Delta\tau\le0.1$ are sufficient for an accurate description of the ground-state properties of the SU(3) Hubbard model.  $\beta=\frac{10}{3}L$ is adopted for calculating unequal-time correlations. Typically, the time complexity of the QMC method for the SU(2) Hubbard model is about $\frac{\beta}{\Delta\tau}\left(2L^2\right)^3$. However, the time complexity of our method is increased to $\frac{\beta}{\Delta\tau}\left(6L^2\right)^3$. The auxiliary field and the time complexity of the QMC method are explained in detail in Supplymentary Information.

%%%%%%%%%%%%%%%%%%%%%%%%%%%%%%%%%%%%%%%
\section{Trion formation}
For studying the trion formation, we consider the correlation function \cite{Kantian2009,Molina2009},
\begin{equation}
    T(i,j) = \langle{ n_{i1}n_{j2}n_{j3} }\rangle,
\end{equation}
which measures the correlation between the color-$1$ fermion at site $i$ and the color-$2$ and color-$3$ fermions at site $j$.
Then the probability of triple occupancy can be defined as
\begin{equation}
    P(3)=\frac{1}{2L^2}\sum_i T(i,i),
\end{equation}
where $L$ is the lattice size of a honeycomb lattice. At strong coupling, $P(3)$ corresponds to the occupancy probability of on-site trion.
%Due to the particle-hole symmetry at half filling, the probability functions, $P(n)$ with $n$ the particle number, satisfy $P(0)=P(3)$, $P(1)=P(2)$ and $P(2)+P(3)=\frac{1}{2}$.

\begin{figure}[tb]
\includegraphics[width=0.96\linewidth]{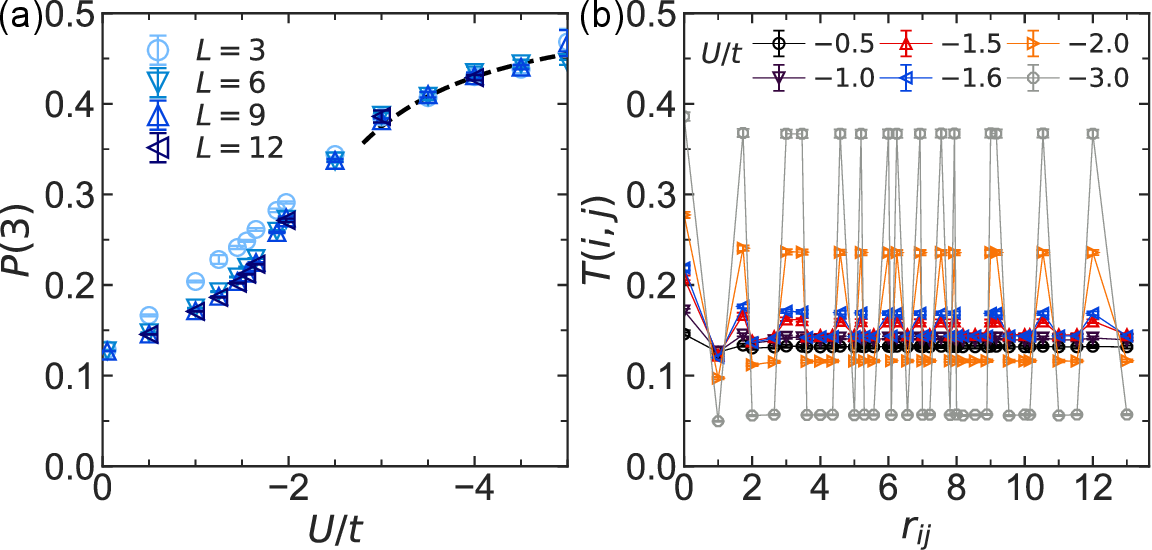}
  \caption{(a)~Probability of the triple occupancy as a function of $U$ for lattice sizes $L=3,6,9$ and $12$. The black dashed curve is the plot of Eq.~\eqref{eq:p3on}. (b)~Correlation function $T(i,j)$ as a function of the distance $r_{ij}\equiv|{\bm r_{i}-\bm r_{j}}|$. The lattice size $L=9$.
}
\label{fig:Tij}
\end{figure}

In the noninteracting limit, the density-density correlation can be decoupled directly, and thus $T(i,i)$ follows the binomial distribution, $\lim_{U/t\to 0}P(3)=\frac{1}{8}$. In the atomic limit, since only on-site trions exist, each site is either fully occupied or empty, and then $\lim_{t/U\to 0}P(3)=\frac{1}{2}$.
In Fig.~\ref{fig:Tij}(a), $P(3)$ is plotted as a function of $U$ for various lattice sizes $L$.
For $L=9$ and $12$, the $P(3)$ versus $U$ curves are almost indistinguishable, and therefore the lattice size $L\geqslant 9$ is sufficiently large to estimate the $L\to\infty$ limit of $P(3)$. As expected, $P(3)$ increases monotonically with $U$.

As shown in Fig.~\ref{fig:Tij}(a), at strong coupling triple occupancy probability $P(3)$ is observably smaller than the large-$U$ limit $\frac{1}{2}$, which infers that there may exist another type of three-body bound state besides the on-site trion in the strong coupling regime.
To explore the possible trion states, let us first consider a two-site half-filled attractive SU(3) Hubbard model.
In this two-site model, there are obviously two possible trion states: on-site trion $\left|\Phi_\mathrm{t}\right\rangle= |123\rangle$ and off-site trion $\left|\Phi_\mathrm{ot}\right\rangle= \frac{1}{\sqrt{3}}\sum_{\cal P}\epsilon_{\cal P} \mathcal{P}|12,3\rangle$,
where the permutation $\mathcal{P}\in \lbrace(1),(13),(23)\rbrace$ and $\epsilon_{\cal P}=(-1)^{\cal P}$ is the parity.
By the first-order perturbation theory, we obtain the ground-state wave function,
\begin{equation}\label{eq:twosite-wf}
    \left|{\Psi_\text{2-site}}\right\rangle = \left|\Phi_\mathrm{t}\right\rangle - \frac{\sqrt{3}}{2}\frac{t}{U}\left|\Phi_\mathrm{ot}\right\rangle,\  t/U\to 0.
\end{equation}
This suggests that the ground state is the superposition of the on-site trion state $\left|\Phi_\mathrm{t}\right\rangle$ and the off-site trion state $\left|\Phi_\mathrm{ot}\right\rangle$. We may use this result to estimate the probability of triple occupancy on the honeycomb lattice:
\begin{equation}\label{eq:p3on}
    P(3)=\frac{1}{2}-\frac{3z}{8}\frac{t^2}{U^2},\  t/U\to 0.
\end{equation}
where the coordination number $z=3$.
For $|{U}|/t\geqslant 3$, the values of $P(3)$ obtained by QMC simulations can be fitted into the equation $P(3)=\frac{1}{2}-a\frac{t^2}{U^2}$.
Surprisingly, the fitting coefficient $a=1.10(3)$ quantitatively agrees with Eq.~\eqref{eq:p3on}.

In Fig.~\ref{fig:Tij}(b), the correlation function $T(i,j)$ is plotted as a function of $r_{ij}$.
Here $r_{ij}\equiv|{\bm r_{i}-\bm r_{j}}|$ is the distance between sites $i$ and $j$ and the distance between nearest-neighbor sites is set as the length unit.
For various $U$, the maximum of each $T(i,j)$ appears at $r_{ij}=0$, due to the attractive interaction between fermions, while the minimum of each $T(i,j)$ appears at $r_{ij}=1$, optimizing the kinetic energy gain that results in an effective nearest-neighbor repulsive interaction between on-site trions at strong coupling \cite{Klingschat2010,Titvinidze2011}.
The minimum of $T(i,j)$ decreases with the increase of $|U|$ and vanishes in the strong coupling limit, which is consistent with the off-site trion term in Eq.~\eqref{eq:twosite-wf}.
Additionally, the interaction-induced CDW phase transition can be illustrated in terms of the behavior of $T(i,j)$.
For $|U|/t\leqslant 1$, $T(i,j)$ converges to a constant when $r_{ij}$ is large, so the correlation function has the equal value for the color-$1$ fermion occupying the two sublattices, reflecting the lattice inversion symmetry of the semimetal phase.
In contrast, for $|U|/t\geqslant 2$, $T(i,j)$ defined on the same sublattice is much larger than that defined on different sublattices, which corresponds to the lattice inversion symmetry breaking of the CDW phase.

%-----------------------------------------------------------
%-----------------------------------------------------------

%-----------------------------------------------------------
\begin{figure}[tb]
  \includegraphics[width=0.9\linewidth]{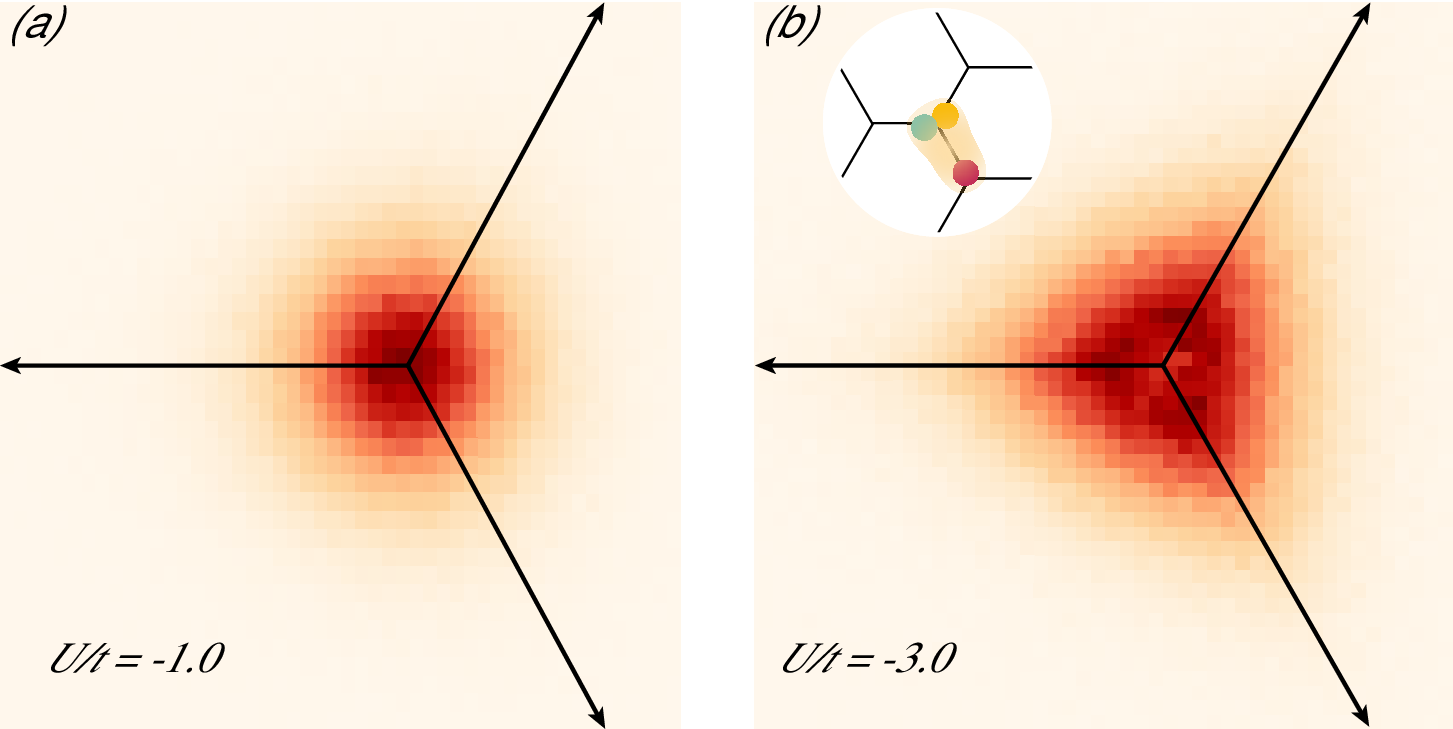}
  \caption{Normalized bond vector histograms $P(N_x,N_y)$ obtained by QMC simulations. The three bond orientations $\hat{e}_a$ are denoted by the long arrows. (a) $U/t=-1.0$ in the semimetal phase. (b) $U/t=-3.0$ in the CDW phase. The lattice size $L=9$.}
  \label{fig:hist2d}
\end{figure}
%-----------------------------------------------------------

The off-site trion at site $i$ can be demonstrated via the probability distribution $P(N_x,N_y)$ of the bond vector $(N_x,N_y) = \sum_{a=1}^{3} \langle d_{i,\hat{e}_a} \rangle_{s} \hat{e}_a$, where $d_{i,\hat{e}_a} = \sum_{\alpha=1}^{3} (tc_{i\alpha}^{\dagger}c_{i+\hat{e}_a\alpha} + \mathrm{H.c.}) $ is the kinetic bond operator; $\langle~\rangle_s$ represents the simulated value during a QMC run; $\hat{e}_a$ denotes three nearest-neighbor bond orientations \cite{albuquerque2011phase,lang2013dimerized,Zhou2016,Zhou2017Finite}.
For a reference lattice point, each of the local bond vectors arising over the QMC simulation is tracked and collected. The entire collection of simulated bond vectors is plotted as a histogram which visualizes the probability distribution of the bond vectors in terms of density of data points.
%In the noninteracting limit, the kinetic bond vectors $\langle d_{i,\hat{e}_a} \rangle \hat{e}_a$ of free fermions along the three orientations are equal and cancel each other out after summing over $a$. Fluctuations in QMC simulations certainly weaken the
%magnitudes of $\langle d_{i,\hat{e}_a}\rangle$, so $P(N_x,N_y)$ are symmetric around the origin.
In the semimetal phase, the probability distribution $P(N_x,N_y)$ are symmetric around the origin since the bond vector $\sum_{a=1}^{3} \langle d_{i,\hat{e}_a}  \rangle_{s} \hat{e}_a$ is homogenous.
In the CDW phase, the off-site trion causes a nonzero $\langle d_{i,\hat{e}_a} \rangle_{s}$. Since there is only one off-site trion on each site, $\sum_{a} \langle d_{i,\hat{e}_a} \rangle_{s} \hat{e}_a$ is polarized along one $\hat{e}_a$ direction.
Figure~\ref{fig:hist2d}(a) shows the symmetric probability distribution $P(N_x,N_y)$ at $U/t=-1$, which indicates the absence of an off-site trion.
As shown in Fig.~\ref{fig:hist2d}(b), at $U/t=-3$, the dominant weight of $P(N_x,N_y)$ is polarized along the three $\hat{e}_a$ directions, manifesting the bond state of an off-site trion.
The histogram only illustrates the formation of a local off-site trion bond state, and does not imply the emergence of a long-range order with off-site trions due to the low density of off-site trions.

%%%%%%%%%%%%%%%%%%%%%%%%%%%%%%%%%%%%%%%
\section{The CDW phase transition}
The CDW ordering can be characterized by the CDW structure factor, which is defined at the $\mathbf{\Gamma}$ point via the density-density correlation function $C(i,j) = \sum_{\alpha,\beta}\langle{ n_{i\alpha} n_{j\beta} }\rangle$\cite{Lee2009},
\begin{equation}
    S_{\mathrm{CDW}}(L,\mathbf{\Gamma})=\frac{1}{2L^2}\sum_{ij}C(i,j)\varepsilon^{i}\varepsilon^{j},
\end{equation}
where $\varepsilon^i=+1$ for sublattice $A$ and $\varepsilon^i=-1$ for sublattice $B$.
Then the CDW order parameter is given by $D = \lim_{L\rightarrow\infty} \sqrt{\frac{1}{2L^2}S_{\mathrm{CDW}}(L,\mathbf{\Gamma})}$.

\begin{figure}[tb]
    \centering
    \includegraphics[width=0.96\linewidth]{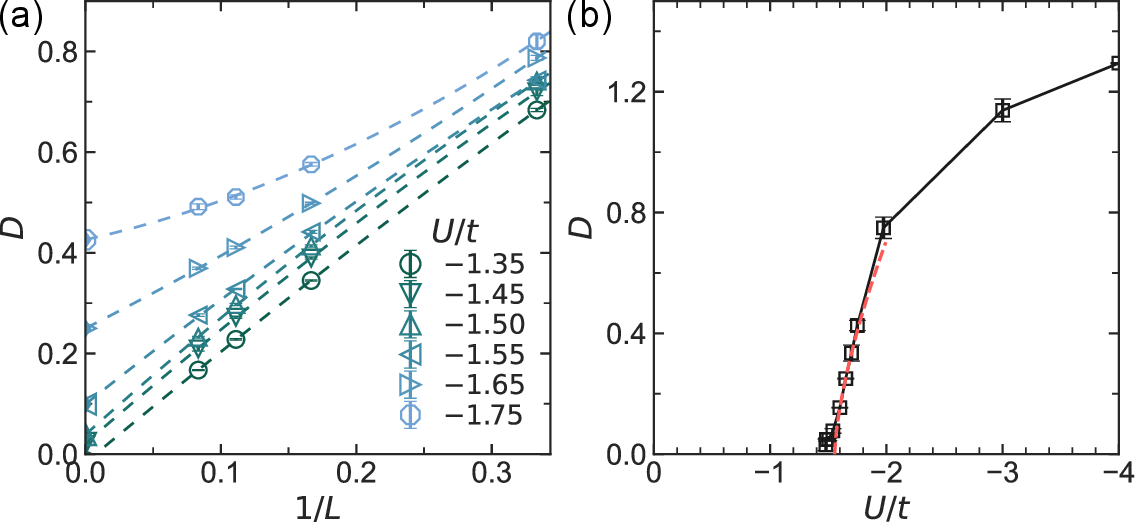}
    \caption{(a)~Extrapolation of the CDW order parameter to the $L\to\infty$ limit for various $U$. The quadratic polynomial fitting is used.
    (b)~Plot of the extrapolated CDW order parameter as a function of $U$. The red dashed curve fits the data to $D\sim |U-U_c|^{-\zeta}$.
    \label{fig:cdworder}
    }
\end{figure}

In the $L\to\infty$ limit, the CDW order parameter for various $U$ can be obtained by finite-size extrapolation.
Figure~\ref{fig:cdworder}(a) shows that the extrapolated CDW order parameters are substantially greater than zero when $|U|/t>1.5$, suggesting that the critical point is at around $U/t \approx -1.5$. The CDW order parameters near the critical point obey the power law  $D\sim|U-U_c|^{-\zeta}$ \cite{Fisher1989Boson,Otsuka2018}, which in turn can be used to fit the extrapolated CDW order parameters and then the critical exponent $\zeta=-0.67(3)$ and the critical coupling $U_c/t=-1.55(1)$ can be extracted, as shown in Fig.~\ref{fig:cdworder}(b). Note that the few non-vanishing extrapolated values below critical point may originate from the simple finite-size extrapolation method and should be removed by the curve-fitting approach \cite{Assaad2013,Otsuka2018}. Compared to the mean-field critical point $U^{\mathrm{MF}}_c/t\approx-1.11$ (see Appendix~\ref{app:MF}), it is reasonable that  $|{U^\mathrm{MF}_c}|<|{U_c}|$ , because quantum fluctuations are neglected in the mean-field approach.

%-----------------------------------------------------------
%-----------------------------------------------------------

%-----------------------------------------------------------
\begin{figure}[tb]
    \includegraphics[width=0.96\linewidth]{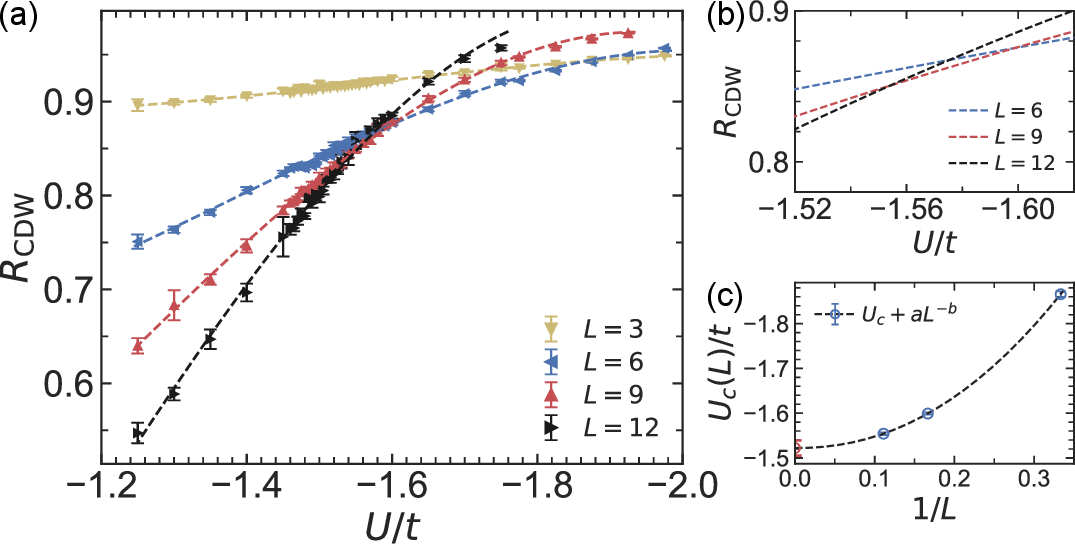}
    \caption{(a) The CDW correlation ratio as a function of $U$ for lattice sizes $L=3,6,9$ and $12$.
    (b) A zoom-in view of the curve-crossing region for lattice sizes $L=6,9$ and $12$.
    (c) The finite-size extrapolation of the crossing points with the error bars from resampling.
    }
    \label{fig:cdwratio}
\end{figure}
%-----------------------------------------------------------

To locate the transition point more accurately, we consider the dimensionless correlation ratio \cite{Chen2019},
\begin{equation}
    R_{\mathrm{CDW}} =1- \frac{S_{\mathrm{CDW}}(L,\mathbf{\Gamma}+\delta\bm{k})}{S_{\mathrm{CDW}}(L,\mathbf{\Gamma})},
\end{equation}
where $\mathbf{\Gamma}$ is the CDW wave vector and $\mathbf{\Gamma}+\delta \bm{k}$ represents a neighboring wave vector in the reciprocal lattice.
In the semimetal phase, $R_{\mathrm{CDW}}$ tends to zero as $S_{\mathrm{CDW}}(L,\mathbf{\Gamma})\approx S_{\mathrm{CDW}}(L,\mathbf{\Gamma}+\delta \bm{k})$.
When the CDW order develops, $S_{\mathrm{CDW}}(L,\mathbf{\Gamma})\gg S_{\mathrm{CDW}}(L,\mathbf{\Gamma}+\delta \bm{k})$, so $R_{\mathrm{CDW}}$ approaches unity.
For sufficiently large $L$, $R_{\mathrm{CDW}}$ curves intersect at size-independent point corresponding to the critical point $U_c$.
For finite lattice sizes, the crossing point of $R_{\mathrm{CDW}}$ curves defines a finite-size estimate of the critical value $U_c(L)$, which takes the form of $U_c(L)=U_c+aL^{-b}$, when taking account of scaling corrections \cite{Parisen2015}.
The critical point $U_c$ is then extracted in the $L\to\infty$ limit.

We use the resampling method \cite{efron1994introduction,Weber2018Two} to extract the crossing points $U_c(L)$ between the $R_{\mathrm{CDW}}(L)$ and $R_{\mathrm{CDW}}(L+3)$ curves.
In Figs.~\ref{fig:cdwratio}(a) and \ref{fig:cdwratio}(b), $R_{\mathrm{CDW}}$ curves show a size-dependent crossing point inbetween $-2.0<U/t<-1.5$ due to significant finite-size effects.
Nevertheless, as shown in Fig.~\ref{fig:cdwratio}(c), the crossing points are fitted into the curve equation $U_c(L)=U_c+aL^{-b}$ where the fitting parameters are $a=-3.7(11)$ and $b=2.2(3)$, and the critical point is found to be $U_c/t=-1.52(2)$.

%-----------------------------------------------------------
\begin{figure}[tb]
    \includegraphics[width=0.96\linewidth]{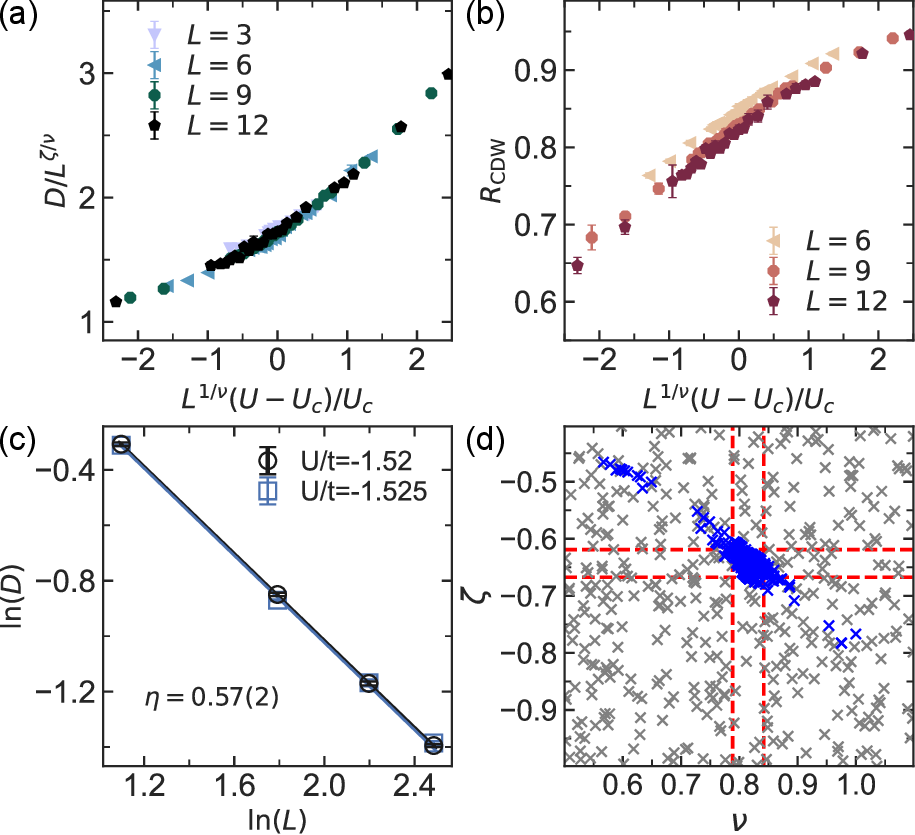}
    \caption{Scaling collapses of (a) the CDW order parameters and (b) correlation ratios by using the exponents $\nu=0.82$, $\zeta=-0.64$.
    (c) The log-log plot of order parameter versus $L$ in the vicinity of critical point.
    (d) Best-fitting analysis of the critical exponents. The converged values are blued while the initial guess values are greyed. The red dashed lines represent the standard errors.
    The critical point $U_c/t=-1.52$.
    }
    \label{fig:cdwcollapse}
    \end{figure}
%-----------------------------------------------------------

We shall derive the critical exponents of the semimetal-CDW transition.
In the vicinity of the critical point, the CDW order parameter obeys the scaling equation \cite{WangL2014,Li2015c},
\begin{equation}
    D(\delta u,L)=L^{\frac{\zeta}{\nu}}\widetilde D(\delta uL^{\frac{1}{\nu}}),
    \label{eq:exp_cdw}
\end{equation}
where $\delta u=(U-U_c)/U_c$ and the exponent $\zeta=-\nu(\eta+z)/2$ (setting $z=1$).
%$\eta=-2\frac{\zeta}{\nu}-1$.
Figures~\ref{fig:cdwcollapse}(a) and \ref{fig:cdwcollapse}(b) show the scaling collapses of the CDW order parameters and correlation ratios, respectively.
The exponent $\eta=0.57(2)$ is extracted from the slope of the log-log plot of $D$ versus $L$ at the critical point $U_c/t=-1.52$, as shown in Fig.~\ref{fig:cdwcollapse}(c).
Then in Fig.~\ref{fig:cdwcollapse}(d), the exponents $\eta$ and $\nu$ are extracted simultaneously by using the best-fitting analysis adapted from Refs.~\cite{Melchert2009,*Sorge2015,Houdayer2004}.
Typically, we randomly choose the value of $U_c/t$ at around $-1.52$ and randomly choose the initial guesses of the exponents $\nu$ and $\zeta$, inside a small range.
By the best-fitting procedure, the converged values of the critical exponents are found to be $\nu=0.82(3)$, $\zeta=-0.64(3)$ and $\eta=0.58(4)$.
%The $\chi^2$-evaluations of scaled order parameter are $5.2$ and $2.4$ for $L\geqslant3$ and $L\geqslant6$, respectively.
%Note that $\chi^2$ approaches unity if the data perfectly collapse on a scaling function \cite{Houdayer2004}.
%In addition, the scaled correlation ratio gives larger $\chi^2$ values, $27$ and $2.8$ for $L\geqslant6$ and $L\geqslant9$ respectively, owing to strong finite-size effects.
%In the standard power-law assumption of order parameter, bigger values of $U_c$ and $|{\zeta}|$ are obtained due to the simple extrapolation method.
%To conclude, the critical exponents are $\nu=0.82(3)$ and $\eta=0.58(4)$.

In the framework of the GNY models, the semimetal-CDW transition of SU(3) Dirac fermions belongs to the $N=3$ chiral Ising universality class, of which
the perturbative renormalization-group (RG) calculations suggest the critical exponents $\nu\gtrsim 1$ and $\eta\gtrsim 0.8$ \cite{Ihrig2018}. It is evident that the critical exponents given by GNY models remarkably deviate from our QMC results by $\gtrsim 20\%$.
Surprisingly, our QMC results coincide with the $N=1$ chiral Ising universality class, of which the perturbative RG calculations give the critical exponents $\nu=0.898(27)$ and $\eta=0.487(12)$ \cite{zerf2017four,Ihrig2018}. The functional RG calculations ($\nu=0.930(4)$, $\eta=0.5506$) \cite{Knorr2016} and the QMC simulations ($\nu=0.88(2)$, $\eta=0.54(6)$) \cite{Huffman2017} of the $N=1$ chiral Ising universality class also suggest similar critical exponents.

%%%%%%%%%%%%%%%%%%%%%%%%%%%%%%%%%%%%%%%
\section{Conclusions and outlook}
%\label{conclusion}
We have performed the sign-problem-free QMC simulations to investigate the trion formation and quantum phase transition in the half-filled attractive SU(3) Hubbard model on a honeycomb lattice. With increasing attractive Hubbard interaction, the continuous semimetal-CDW transition occurs at the critical point $U_c/t=-1.52(2)$ and the corresponding critical exponents are $\nu=0.82(3)$ and $\eta=0.58(4)$. In the CDW region, the off-site trions emerge due to density fluctuations, and therefore the on-site and off-site trions coexist in the deep CDW phase.

Our QMC simulations illustrate the formation of a local off-site trion bond state in two-dimensional Hubbard model, which extends the understanding of one-dimensional off-site trions suggested by previous DMRG study.
It has been proposed that the trionic phase can be probed by shaking the optical lattice \cite{Rapp2007}. Moreover, the triple occupancy can be determined experimentally by measuring the loss of atoms governed by a three-body process \cite{Ottenstein2008}.
Our work opens a new avenue for exploring the physical effects of the interplay between on-site and off-site trions in two-dimensional spatial models.

What is particularly interesting about our findings is that the critical exponents determined by QMC simulations remarkably disagree with those of the $N=3$ chiral Ising universality class predicted by the effective GNY theory which has been believed to suggest a general description for the criticality of two-dimensional Dirac fermions and has been numerically verified in several SU($N$) models. Unexpectedly, our QMC results are in fair agreement with the $N = 1$ chiral Ising universality class. We argue that the formation of trions may affect the quantum criticality of attractive SU($3$) Dirac fermions. At the critical point, two species of fermions (unbound fermions and trions) get involved in the ongoing development of CDW order. The on-site trion as a whole can be recognized as a spinless fermion, and dominate the long-range CDW ordering at strong coupling. Within the framework of GNY model, the criticality of spinless trions belongs to the $N = 1$ chiral Ising universality class. Therefore formation of trions completely deprive the attractive SU($3$) Hubbard model of the $N = 3$ chiral Ising universality class. However, the critical point at which unbound fermions are still in the majority is far from the strong coupling regime. The reason that the QMC results are in a good coincidence with the $N = 1$ chiral Ising universality class is still uncertain. Our results evidently raise doubts about the extent of applicability of GNY model on attractive SU($3$) Dirac fermions, which is definitely worth to think and need to solve in future study.

%-----------------------------------------------------------
\acknowledgments
This work is financially supported by the National
Natural Science Foundation of China under Grants No. 11874292, No. 11729402, and No. 11574238. X. W. acknowledges the support from Research Grants Council of the Hong Kong Special Administrative Region, China (No. CityU 11303617), the National Natural Science Foundation of China (No. 11874312) and the Guangdong Innovative and Entrepreneurial Research Team Program (No. 2016ZT06D348). L. W. is supported by the Ministry of Science and Technology of China under the Grants No. 2016YFA0300603 and No. 2016YFA0302400.
C.W. is supported by the National Natural Science Foundation of China under the Grants No. 12174317
and No. 12234016. We acknowledge the support of the Supercomputing Center
of Wuhan University.

%-----------------------------------------------------------

% APPENDIX
\appendix

\appsection{Estimation of the Mott gap}
\label{app:largeUgap}

The attractive SU($N$) Hubbard Hamiltonian consists of two parts:
$H=H_U+H_0$ with $H_U=U\sum_{i,\alpha<\beta}(n_{i\alpha}-1/2)(n_{i\beta}-1/2)$ and $H_0=-t\sum_{\langle{ij}\rangle,\alpha}(c^{\dagger}_{i\alpha}c_{j\alpha}+\mathrm{H.c.})$.
At half-filling, we define the Mott-insulating state by the nonzero energy penalty of adding (removing) a particle into (from) the system \cite{Zhou2014,Zhou2016}.
In the atomic limit the energy penalty of adding an extra hole into the Mott-insulating state is $\Delta_\mathrm{at}=E_{N_{\rm tot}-1}-E_{N_{\rm tot}}=-\frac{U}{2}(N-1)$ where $N_{\rm tot}$ is the total number of fermions.
When tuning on the hopping term, the extra hole (particle) can move in the system, which further modifies the energy penalty.
Below, we derive an effective model for the description of the extra hole (particle).

\begin{figure}
  \includegraphics[width=0.9\linewidth]{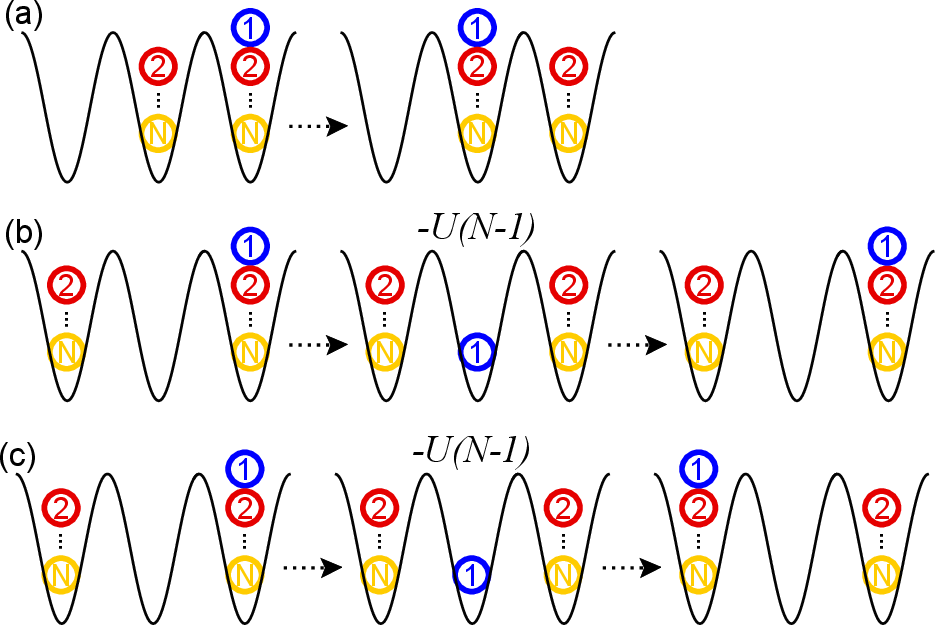}
  \caption{In the attractive SU($N$) Hubbard model, (a) the first-order process inside the subspace $\cal D$ and (b)(c) the second-order processes with the energy penalty $-U(N-1)$.}
  \label{fig:supp:atomic_1d}
\end{figure}

Following the steps in Ref.\cite{sakurai1995modern}, we define the ground state $|n^{(0)}\rangle$ by $ H_U|n^{(0)}\rangle=E_D^{(0)}|n^{(0)}\rangle$ with $E_D^{(0)}=E_{N_{\rm tot}-1}$.
Let $P_0$ be the projection operator onto the subspace $\cal D$ spanned by $|n^{(0)}\rangle$. The projection operator outside the subspace $\cal D$ is then defined as $ P_1=1- P_0$.
The degenerate Rayleigh-Schr{\"o}dinger perturbation theory yields the effective Hamiltonian, $ H_{\rm eff}|n^{(0)}\rangle=E_n|n^{(0)}\rangle$, up to the second order,
\begin{equation}
   H_{\rm eff}= P_0\left(E_D^{(0)}+ H_0+ H_0 P_1\frac{1}{E_D^{(0)}- H_U} P_1 H_0\right) P_0,
\end{equation}
where the hopping operator $H_0P_0$ moves one fermion from the local singlet bound state to the NN sites, as shown in Fig.~\ref{fig:supp:atomic_1d}.
Consider the Pauli exclusion principle, the Hamiltonian can be written as
\begin{eqnarray}\label{eq:supp:heff_hole}
  \nonumber &\begin{aligned}
      H_{\rm eff}=&E_D^{(0)}+C^{(1)}+C^{(2)} \\
          &+\frac{t^2}{U(N-1)} P_0\sum_{\langle\langle{ij}\rangle\rangle\in{A},\alpha}(c^{\dagger}_{i\alpha}c_{j\alpha} +{\rm H.c.}) P_0,\\
      =&E_{N_{\rm tot}-1}+C^{(1)}+C^{(2)}\\
          &+\frac{t^2}{U(N-1)}\sum_{\langle\langle{ij}\rangle\rangle\in{A}}(c^{\dagger}_{i1}c_{j1} +{\rm H.c.}),\\
  \end{aligned} \\
  &=E_{N_{\rm tot}-1}+C^{(1)}+C^{(2)}+\sum_{\bm k}\epsilon_{\rm h}(\bm{k})c^{\dagger}_{\bm k1}c_{\bm k1},
\end{eqnarray}
where $\langle\langle{ij}\rangle\rangle$ represents the next-nearest-neighbor (NNN) sites and the NNN hopping process is restricted to the fermion color of the extra hole.
$C^{(1)}= P_0 H_0 P_0$ is the first-order perturbation (see Fig.~\ref{fig:supp:atomic_1d}(a)), and
$C^{(2)}$ is the second-order perturbation where one fermion hops to the NN sites and then hops back (see Fig.~\ref{fig:supp:atomic_1d}(b)).
The last term represents the effective tight-binding model originated from the desired second-order process, as shown in Fig.~\ref{fig:supp:atomic_1d}(c).
In the ground state, we assume that the extra hole (particle) occupies the lowest energy level $-y t'$, where $t'=-\frac{t^2}{U(N-1)}$ and $y>0$ is a constant relying on the lattice structure.

Similarly, the effective Hamiltonian of the half-filled system yields
$H_{\rm eff}=E_{N_{\rm tot}}+{C}^{(2)}$
where the NNN hopping process is absent up to the second order.
Combined with Eq.~\eqref{eq:supp:heff_hole}, we obtain the energy penalty
\begin{equation}
  \Delta_\mathrm{at}=-\frac{U}{2}(N-1)+\frac{y t^2}{U(N-1)}+C^{(1)}+\delta C^{(2)},
\end{equation}
where $\delta C^{(2)}$ is the difference between the NN second-order correction terms. In fact, $\delta C^{(2)}$ is marginal and we may also safely discard $C^{(1)}$ because of the CDW ordering. At the Mott transition point, $\Delta_\mathrm{at}=0$ and then an estimate of the transition point is
\begin{equation}
  U^\mathrm{at}_c=-\frac{\sqrt{2y }t}{N-1}.
  \label{eq:supp:uc2nd}
\end{equation}

For the attractive SU($N$) Hubbard model on the honeycomb lattice,
the NNN hopping processes are on a triangular lattice as shown in Fig.~\ref{fig:mottgap}, and thus $\epsilon_{\rm h}(\bm{k})=-2t'\sum_{i=1}^{3}\cos(\bm{k}\cdot\hat{\delta}_i)$
with the primitive vectors $\hat{\delta}_{1}=(\sqrt{3},0)$, $\hat{\delta}_{2}=(\frac{\sqrt{3}}{2},\frac{3}{2})$ and $\hat{\delta}_{3}=(\frac{\sqrt{3}}{2},-\frac{3}{2})$.
Thus, the lowest energy level is $\epsilon_{\rm h}(\mathbf{\Gamma})=-6t'$.
Substituting $y =6$ into Eq.~\eqref{eq:supp:uc2nd},
we obtain
\begin{equation}
  U^\mathrm{at}_c/t=-\frac{2\sqrt{3}t}{N-1}.
\end{equation}
For $N=3$, the critical point is estimated as $U^\mathrm{at}_c/t=-\sqrt{3}$.

%===================================================
%===================================================

\appsection{The auxiliary fields coupled to on-site color flips}
\label{app:HS}
In this section, we adopt the theorem from Ref.~\cite{WangL2015} to prove that the QMC simulation of attractive SU(3) Hubbard model is sign-problem-free at half-filling.
We consider the finite-temperature formalism and apply the Suzuki-Trotter decomposition to separate the kinetic and interaction terms in the partition function,
\begin{equation}
    Z=\mathrm{tr}\left\{e^{-\beta  H}\right\} = \mathrm{tr}\left\{\left[\prod_{k=1}^M e^{-\Delta\tau  H_0}e^{-\Delta\tau  H_U}\right]\right\},%+{\cal O}\left(\Delta\tau^2\right),
    \label{eq:pf}
\end{equation}
where $\Delta\tau=\frac{\beta}{M}$ is the Trotter decomposition step. For the attractive Hubbard interaction,
\begin{equation}
    \begin{aligned}
        &e^{-\Delta\tau U \sum_{\alpha<\beta}\left(n_{\alpha}-\frac{1}{2}\right) \left(n_{\beta}-\frac{1}{2}\right)}\\
        &=\prod_{\alpha<\beta}e^{-\frac{\Delta\tau U}{2}\left( c^{\dagger}_{\alpha}c_{\beta}-\mathrm{H.c.}\right)^2-\frac{\Delta\tau U}{4}}.
    \end{aligned}
    \label{eq:ffip}
\end{equation}
The coupling matrix between colors $\alpha$ and $\beta$ reads
\begin{equation}
    \big(c^{\dagger}_{\alpha},c^{\dagger}_{\beta}\big)
    \left(\begin{array}{cc}
        0 & 1\\
        -1 & 0
    \end{array}\right)
    \left(\begin{array}{c}
        c_{\alpha}\\
        c_{\beta}
    \end{array}\right),
    \label{eq:flmat}
\end{equation}
and its eigenvectors correspond to the complex fermion basis \cite{Li2015b},
\begin{equation}
    \tilde c_{\alpha\beta}=\frac{1}{\sqrt{2}}\left(c_{\alpha}-ic_{\beta}\right),\
    \tilde c_{\beta\alpha}=\frac{1}{\sqrt{2}}\left(c_{\alpha}+ic_{\beta}\right).
\label{eq:cbasis}
\end{equation}

It can be verified that $\tilde c_{\alpha\beta}$ and $\tilde c_{\beta\alpha}$ obey fermionic anticommutation relations:
$\lbrace \tilde c^{\dagger}_{\alpha\beta},\tilde c_{\alpha\beta}\rbrace=1,~\lbrace \tilde c^{\dagger}_{\alpha\beta},\tilde c_{\beta\alpha}\rbrace=0$.
After the diagonalization of coupling matrix Eq.~\eqref{eq:flmat}, the Hubbard interaction term
$(c^{\dagger}_{\alpha}c_{\beta}-c^{\dagger}_{\beta}c_{\alpha})^2$ %in Eq.~\eqref{eq:ffip}
is refined to
$-( \tilde c^{\dagger}_{\alpha\beta}\tilde c_{\alpha\beta}-\tilde c^{\dagger}_{\beta\alpha}\tilde c_{\beta\alpha})^2$.
In the subspace spanned by the complex fermion basis, we apply the discrete Hubbard-Stratonovich (HS) transformation to decouple the Hubbard interaction term:
\begin{equation}
    e^{\frac{\Delta\tau U}{2}\left( \tilde c^{\dagger}_{\alpha\beta}\tilde c_{\alpha\beta}-\tilde c^{\dagger}_{\beta\alpha}\tilde c_{\beta\alpha}\right)^2}
    =\frac{1}{2}\sum_{s_{\alpha\beta}=\pm 1} e^{is_{\alpha\beta}\lambda\left(\tilde c^{\dagger}_{\alpha\beta}\tilde c_{\alpha\beta}-\tilde c^{\dagger}_{\beta\alpha}\tilde c_{\beta\alpha}\right)},
\label{eq:cHS}
\end{equation}
where $\lambda=\arccos{e^{\frac{\Delta\tau U}{2}}}$ and $U<0$.
From Eqs. \eqref{eq:cbasis} and \eqref{eq:cHS}, we may write Eq.~\eqref{eq:ffip} as
\begin{equation}
    \begin{aligned}
    &e^{-\Delta\tau U\left(n_{\alpha}-\frac{1}{2}\right)\left(n_{\beta}-\frac{1}{2}\right)}\\
    &=\frac{1}{2}e^{-\frac{\Delta\tau U}{4}}\sum_{s_{\alpha\beta}=\pm 1} e^{s_{\alpha\beta}\lambda\left( c^{\dagger}_{\alpha}c_{\beta}-c^{\dagger}_{\beta}c_{\alpha}\right)}.
    \end{aligned}
\end{equation}
%In order to have an effective Hamiltonian in the explicit form of anti-symmetric matrix on the exponential,

\begin{figure}[tb]
    \includegraphics[width=0.96\linewidth]{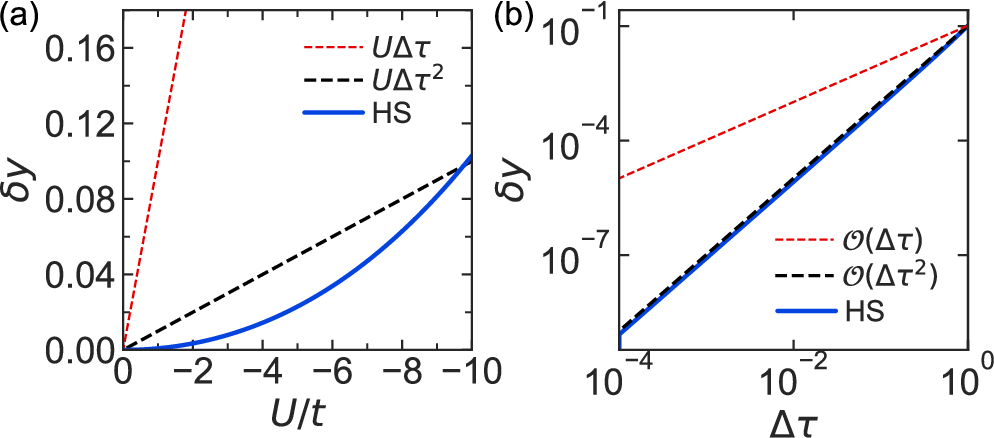}
    \caption{Correctness test of the HS transformation: (a) The numerical error as a function of $U$ for $\Delta\tau=0.1$; (b) The numerical error as a function of $\Delta\tau$ for $U/t=-1$. The red and black dashed curves are respectively the Trotter errors of $\Delta\tau$ and $\Delta\tau^2$.
    }
    \label{fig:supp:hs}
    \end{figure}

It is seen that the Hubbard interaction term in the SU(3) Hamiltonian Eq. \eqref{eq:hubbard} can be decoupled in the on-site color-flip channels via three auxiliary fields $s_{12}, s_{13}$ and $s_{23}$:
\begin{equation}
    \begin{split}
    &e^{-\Delta\tau U\sum_{\alpha<\beta}\left(n_{\alpha}-\frac{1}{2}\right)\left(n_{\beta}-\frac{1}{2}\right)}\\
    &=\frac{1}{8}e^{-\frac{3\Delta\tau U}{4}}\sum_{s_{12},s_{13},s_{23}=\pm 1} e^{\sum_{\alpha<\beta}s_{\alpha\beta}\lambda\left( c^{\dagger}_{\alpha}c_{\beta}-c^{\dagger}_{\beta}c_{\alpha}\right)},
    \end{split}
    \label{eq:hs}
\end{equation}
with a systematic error ${\cal O}(\Delta\tau^2)$.
We expand Eq.~\eqref{eq:hs} in the \textit{on-site} Fock space,
and plot the numerical error of the diagonal element as a function of $U$, as shown in Fig.~\ref{fig:supp:hs}.
According to this correctness test, the HS transformation used in our work is accurate in the intermediate coupling regime ($U/t<10$) but debatable at very strong couplings ($U/t>10$).

%===================================================
%===================================================

Next, we prove that QMC simulations using the HS transformation of Eq.~\eqref{eq:hs} can avoid the sign problem. In the bipartite lattice, we arrange the color-orbital operators in such order:
  $(A_{\alpha},A_{\beta},A_{\gamma},B_{\alpha},B_{\beta},B_{\gamma})$,
where $A$ and $B$ represent the two sublattices. Then, the Hamiltonian after the HS transformation takes the form of
    $H_s=\left(\begin{array}{cc}D_A & K \\ K & D_B\end{array}\right)$
where $K$ represents the hopping matrix between two sublattices and $D_{A(B)}$ is a block diagonal matrix.
In particular, each block in $D_{A(B)}$ is an anti-symmetric matrix,
\begin{equation}
    \Lambda=
    \left(\begin{array}{ccc}
        0 & s_{12}\lambda  & s_{12}\lambda \\
        -s_{12}\lambda & 0 & s_{23}\lambda \\
        -s_{13}\lambda & -s_{23}\lambda & 0
    \end{array}\right),
\end{equation}
and $\Lambda=-\Lambda^T$.
It is easy to verify that the decoupled Hamiltonian satisfies
\begin{equation}
  \eta H_s\eta=-H_s^T,
\end{equation}
where
$\eta=\mathbf{diag}({1,\dots,1},{-1,\dots,-1})$.
This condition guarantees the sign-problem-free determinant QMC simulations \cite{WangL2015}.
However, when the system is away from half filling, $D_{A(B)}$ is not a real anti-symmetric matrix, which may cause the sign problem.
Furthermore, we can use the Rodrigues formula to simplify the matrix exponential:
$e^{\Lambda}=I_3+\frac{\sin{\theta}}{\theta}\Lambda+\frac{(1-\cos{\theta})}{\theta^2}\Lambda^2$
with
$\theta=\sqrt{(s_{12}^2+s_{13}^2+s_{23}^2)\lambda^2}=\sqrt{3\lambda^2}$.

The auxiliary fields associated with on-site color-flip channels can be generalized to the attractive SU($2N$+1) Hubbard model.
This HS transformation, however, enlarges the size of matrix which leads to the time complexity ${\cal O}(\beta (2L^2N')^3)$ with $N'=2N+1$.
% Here $N$ is the number of fermion colors and $2L^2$ is the number of sites on a honeycomb lattice.
The HS transformation of Eq.\eqref{eq:hs} can also be applied to the projector QMC (PQMC) method in which the expectation values of observables with the ground state $\left|{\Psi_G}\right\rangle$ are calculated. A trial ground-state wave function $\left|{\Psi_T}\right\rangle$ is chosen based on the Hamiltonian Eq.\eqref{eq:hubbard} in the noninteracting limit under the anti-periodic boundary condition  \cite{WangL2015}.
Given the inner product $\langle{\Psi_G}|{\Psi_T}\rangle\ne 0$,
the ground state $\left|{\Psi_G}\right\rangle$ can be reached by applying the projection operator $e^{-\frac{\beta  H}{2}}$ onto $\left|{\Psi_T}\right\rangle$ for a sufficiently long projection time $\beta$ \cite{Assaad2008,Wang2014}.

In our simulations, the honeycomb lattice in real space is subject to the periodic boundary condition for $L=3,6,9,12$. Under this boundary condition, the lattice vectors in the Brillouin zone can meet the Dirac points.
The projection time $\beta=\frac{8}{3}L$ and the Trotter decomposition step $\Delta\tau\leqslant0.1$ are sufficient for the accurate description of the ground-state properties of the attractive SU(3) Hubbard model.

\begin{figure}[tb]
  \includegraphics[width=0.9\linewidth]{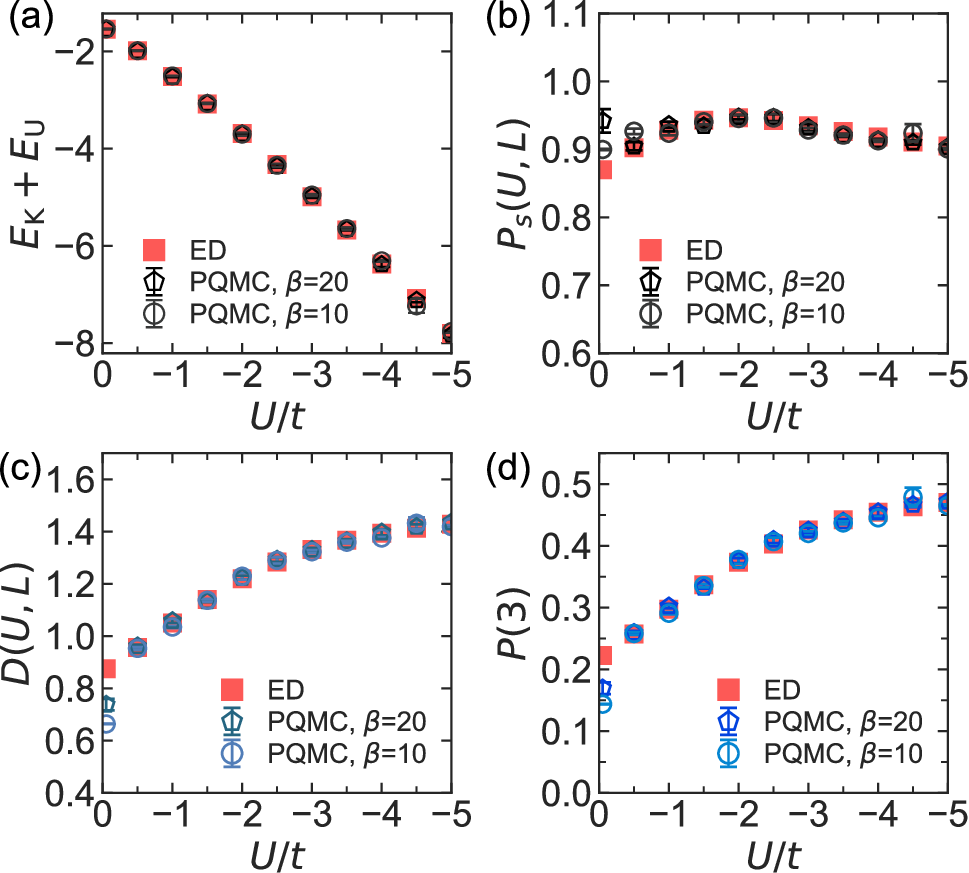}
  \caption{Correctness test on a $2\times2$ square lattice for (a) ground state energy, (b) pairing order parameter, (c) CDW order parameter, and (d) triple occupancy probability.
  }
  \label{fig:supp:ed}
  \end{figure}

In the half-filled attractive SU(3) Hubbard model, the results of PQMC are compared with those of exact diagonalization (ED) method on a $2\times2$ square lattice.
In Fig.~\ref{fig:supp:ed}, the ED results are shown with the red square, while the PQMC results for $\beta=10,20$ are presented respectively by the blue circle and blue pentagon.
For most of $U$, PQMC results agree well with ED calculations within the margin of numerical errors.
However, when $U/t=-0.05$, we obtain inconsistent results except the ground state energy.
The reason could be that the trial wave function becomes inappropriate due to Fermi surface nesting.

%===================================================
%===================================================

\appsection{Mean-field analysis}\label{app:MF}
For the CDW order on a honeycomb lattice,
the Hubbard interaction term can be decoupled in terms of the on-site parameters $\langle{ c^{\dagger}_{i\alpha}c_{i\alpha} }\rangle=\varepsilon^i\Delta$ at the mean-field level,
where $\Delta$ is the CDW order parameter; $\varepsilon^i$ is $+1$ on sublattice $A$ and is $-1$ on sublattice $B$. %, specifying the density imbalance between the two sublattices.
The mean-field CDW order parameter can be defined as
\begin{equation}
        \Delta = \frac{1}{2L^2N}\sum_{i,\alpha}\varepsilon^i\langle{ c^{\dagger}_{i\alpha}c_{i\alpha} }\rangle.
        \label{eq:mf-cdw}
\end{equation}
The mean-field Hamiltonian of the attractive SU(3) Hubbard model is then written in the reciprocal space,
$H_{\mathrm{MF}} = \sum_{\bm{k},\alpha} c^{\dagger}_{\bm{k},\alpha} h_{\bm{k}}c_{\bm{k},\alpha}$, where
\begin{equation}
    h_{\bm{k}} =
    \begin{bmatrix}
    (N-1)U\Delta & \epsilon^{*}(\bm{k}) \\
    \epsilon(\bm{k}) & (1-N)U\Delta
    \end{bmatrix},
\end{equation}
and the bipartite basis $c^{\dagger}_{\bm{k}}\equiv(c^{\dagger}_{A\bm{k}},c^{\dagger}_{B\bm{k}})$ is used. The off-diagonal term $\epsilon(\bm{k})=-t\sum_{a}e^{-i\bm{k}\cdot\hat{e}_a}$ comes from the noninteracting part of Eq.~\eqref{eq:hubbard}, where $\sum_{a}$ is the sum over three vectors $\hat{e}_1=(0,1)$, $\hat{e}_2=(-\frac{\sqrt{3}}{2},-\frac{1}{2})$ and $\hat{e}_3=(\frac{\sqrt{3}}{2},-\frac{1}{2})$.
The distance between nearest-neighbor (NN) sites is set as the unit of length.
At half filling, the self-consistent equation of $\Delta$ reads
\begin{equation}
    \Delta=\frac{1}{2L^2N}\sum_{\bm{k},\alpha}\left(\frac{\epsilon(\bm{k})^2 }{|\lambda_{\bm{k}}|^2+(N-1)U\Delta|\lambda_{\bm{k}}|} - 1\right),
    \label{eq:iter}
\end{equation}
where $\lambda_{\bm{k}}=\pm\sqrt{(N-1)^2U^2\Delta^2+\epsilon(\bm{k})^2}$ are the eigenvalues of the matrix $h_{\bm{k}}$. The nonzero value of $\Delta$ opens the energy gap $2(N-1)|{U}|\Delta$ in the energy spectrum $\lambda_{\bm{k}}$. Thus the ground state is an insulator and Eq.~\eqref{eq:mf-cdw} is often referred to as \textit{the gap function}.
Furthermore, according to Eq.~\eqref{eq:iter} the number of fermion colors actually rescales $U$ by the factor $N-1$ \cite{Koga2017}.

\begin{figure}[tb]
    \includegraphics[width=0.9\linewidth]{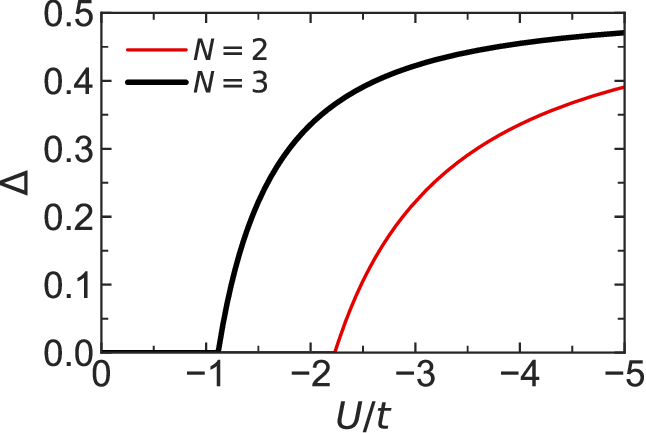}
    \caption{Mean-field solution of the CDW order parameter $\Delta$ as a function of $U$ for $N=2,3$ at half filling.
    }
    \label{fig:supp:mf}
    \end{figure}

For each Hubbard $U$, one can solve Eq.~\eqref{eq:iter} self-consistently by the root-finding method \cite{SciPy2001,Brent2013}. As shown in Fig.~\ref{fig:supp:mf}, the CDW phase transitions occur at the critical points ${U^{\mathrm{MF}}_c}/t\approx-1.11$ for $N=3$ and $U^{\mathrm{MF}}_{c,N=2}/t=2U^{\mathrm{MF}}_c/t\approx-2.23$ for $N=2$.

On the other hand, the pairing gap function is defined as $\Delta_{\alpha\beta}=-\frac{1}{2L^2}\sum_{k}\langle{c_{k\alpha}c_{-k\beta}}\rangle$. The vector $(\Delta_{23},\Delta_{31},\Delta_{12})$ can be mapped onto $(0,0,\Delta_{12})$ by a global gauge change, leaving only one pair gapped \cite{Honerkamp2004BCS}.
By solving the standard BCS problem, the energy spectrum has the gapped branches $\pm\sqrt{\epsilon(\bm{k})^2+U^2\Delta_\mathrm{12}^2}$ with the energy gap $|{2U\Delta_\mathrm{12}}|$.
Note that Eq.~\eqref{eq:iter} also holds for $\Delta_{12}$ when $N=2$. Thus, in the half-filled attractive SU(3) Hubbard model, the pairing gap ($N=2$) is certainly not larger than the CDW gap ($N=3$), and the ground state is associated with CDW order \cite{Honerkamp2004Ultra}.
%This indicates that the critical point in the SU(2) case satisfies $U^{\mathrm{MF}}_{c,N=2}/t=2U^{\mathrm{MF}}_c/t\approx-2.23$ \cite{Sorella1992}.

%===================================================
%===================================================

Next, we investigate the semimetal-CDW transition by analyzing the Ginzburg-Landau (GL) free-energy density $f(\Delta)$.
%We then Fourier transform $\Delta$ as
%\begin{equation}
%\begin{aligned}
%\Delta&=\frac{1}{2L^{2}N}\sum_{\vec{k},\alpha}n_{\vec{k},\alpha,A}-n_{\vec{k},\alpha,B} \\
%&=\frac{1}{2L^{2}N}\sum_{\vec{k},\alpha}\Delta_{\alpha}\left(\vec{k}\right),
%\end{aligned}
%\end{equation}
%in which $n_{\vec{k},\alpha,A}$ and $n_{\vec{k},\alpha,B}$ are the particle number operator for wave vector $\vec{k}$ and color $\alpha$ on sublattice A and B respectively.
Since $\sigma_{x}\Delta\sigma_{x}=-\Delta$ in which $\sigma_{x}$ is the Pauli matrix in the basis of sublattices, the CDW order breaks the lattice inversion symmetry. However, $f(\Delta)$ needs to maintain the lattice inversion symmetry. Hence, the analytic part of $f(\Delta)$ can be written as
\begin{equation}
\label{eq:GL}
f_\mathrm{a}=r_{2}\Delta^{2}+r_{4}\Delta^{4}.
\end{equation}
%where $r_2<0$ and $r_4>0$.
According to the GL theory, $f_\mathrm{a}$ describes a second-order transition with the critical exponent $\zeta=-\frac{1}{2}$.
Due to the coupling between $\Delta$ and the gapless Dirac fermions, the total free-energy density $f(\Delta)$ potentially contains a nonanalytic part as explained below.
When taking account of the spin degeneracy, there are six Dirac cones in the first Brillouin zone.
In the CDW phase, the mean-field energy spectrum around each Dirac point is $E_{k}=\sqrt{v^{2}k^{2}+(N-1)^{2}U^{2}\Delta^{2}}$, where $k$ represents the deviation from the Dirac point.
Denote the nonanalytic part by $f_\mathrm{non}(\Delta,\beta)$ where $\beta$ is the inverse temperature.
At the mean-field level, we can estimate $f_\mathrm{non}(\Delta,\beta)$ arising from the low-energy spectra around the Dirac points as
\begin{equation}
    \begin{aligned}
        f_\mathrm{non}(\Delta,\beta)\approx &
        -\frac{6}{\beta}\int_{0}^{\Lambda}\frac{d^{2}\bm{k}}{(2\pi)^{2}}
        \left[\ln\left(1+e^{\beta E_{k}}\right)\right. \\
        &\left.+\ln\left(1+e^{-\beta E_{k}}\right)\right],
    \end{aligned}
\end{equation}
where $\Lambda$ is the momentum cutoff.
By taking the limit of $\beta\rightarrow \infty$
and then $\Delta \rightarrow 0$,
we solve the integral,
\begin{equation}
f_\mathrm{non}=r_{3} |\Delta|^{3},
\end{equation}
where $r_{3}=\frac{(N-1)^{3}U^{3}}{\pi v^{2}}>0$.
Combined with the analytic part $f_\mathrm{a}$, we obtain the total GL free-energy density
\begin{equation}
f(\Delta)=f_\mathrm{a}+f_\mathrm{non}= r_{2}\Delta^{2}+r_{3}|\Delta|^{3}.
\end{equation}
According to the GL theory, $f(\Delta)$ describes a second-order transition with the critical exponent $\zeta=-1$.

\appsection{Absence of the pairing order}

%===================================================
\begin{figure}[tb]
  \centering
  \includegraphics[width=0.9\linewidth]{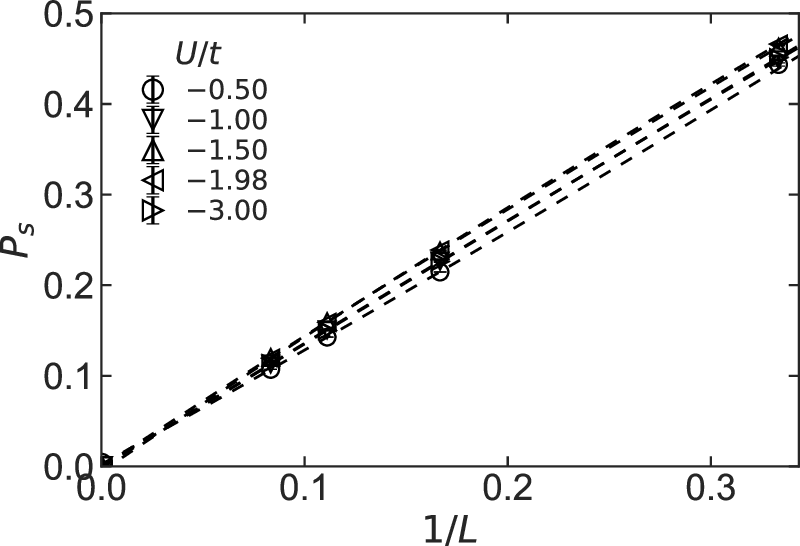}
  \caption{%The pairing order in the attractive SU(3) Hubbard model.
  Finite-size extrapolations of the pairing order parameter to the $L\to\infty$ limit for various $U$. Dashed lines are the least-square fits with the quadratic polynomials in $1/L$.
  \label{fig:supp:pairing}
  }
\end{figure}
%===================================================

\label{sect:upair}
The pairing structure factor can be defined as
\begin{equation}
    S_{\mathrm{pair}}(L)=\frac{1}{2L^2}\sum_{i,j} P(i,j),
\end{equation}
where $2 \times L \times L$ is the number of lattice sites and $P(i,j) = \sum_{\alpha<\beta} \langle{ c^{\dagger}_{i\alpha}c^{\dagger}_{i\beta}c_{j\beta}c_{j\alpha} + c_{i\beta}c_{i\alpha}c^{\dagger}_{j\alpha}c^{\dagger}_{j\beta} }\rangle$ is the equal-time pair-pair correlation function.
By extrapolating the structure factor to thermodynamic limit, the long-range pairing order parameter can be obtained: $P_s = \lim_{L\rightarrow\infty} \sqrt{\frac{1}{2L^2}S_{\mathrm{pair}}(L)}$.

At half filling, the ground state of the attractive SU(3) Hubbard model on the honeycomb lattice is a semimetal in noninteracting limit $U/t\to0$.
As the attractive interaction increases, the system may enter an ordered phase. At weak coupling, the pairing gap function on the honeycomb lattice is vanishingly small because of the zero density of states at Dirac points \cite{Honerkamp2004BCS}, and therefore the quantum fluctuations prevent pairings on the half-filled honeycomb lattice. At strong coupling, the attractive SU(2) Hubbard model enters the ordered phase where the pairing order and the CDW order are degenerate, and thus the (spin) SU(2) symmetry is preserved \cite{Scalettar1989,Moreo1991,Lee2009,Micnas1990super}.

As shown in Fig.~\ref{fig:supp:pairing}, the finite-size extrapolation to the $L\to\infty$ limit shows no sign of pairing order on the honeycomb lattice in a wide range of coupling strengthes. Hence, the pairing order is absent and therefore the color superfluid is not the ground state of our model. In Appendix~\ref{app:MF}, the mean-field analysis also indicates that the energy of pairing order is higher than that of CDW order on a honeycomb lattice, which is consistent with our QMC results.

%%%%%%%%%%%%%%%%%%%%%%%%%%%%%%%%%%%%%%%%%%%%%%%%%%%%%%
%\bibliography{su3_honey.bib}

\end{document}